\documentclass[
reprint,
superscriptaddress,
showpacs,
 amsmath,amssymb,
 aps,
prl,
floatfix,
]{revtex4-1}

\usepackage{graphicx}
\usepackage{dcolumn}
\usepackage{bm}
\usepackage{chemformula}
\usepackage{hyperref}
\usepackage{ulem}

\usepackage{multirow}

\newcommand{\bcgo}{Ba$_2$CoGe$_2$O$_7$}

\begin{document}

\preprint{APS/123-QED}

\title{In-situ electric field control of THz non-reciprocal directional dichroism in the multiferroic \bcgo{}}

\author{J. V\'it}
\affiliation{Department of Physics, Budapest University of Technology and 
	Economics, 1111 Budapest, Hungary}
\affiliation{Institute of Physics ASCR, Na Slovance 2, 182 21 Prague 8, Czech Republic}
\affiliation{Faculty of Nuclear Science and Physical Engineering, Czech Technical University, B\v{r}ehov\'a 7, 115 19 Prague 1, Czech Republic}

\author{J. Viirok}
\author{L. Peedu}
\author{T. R{\~o}{\~o}m}
\author{U. Nagel}
\affiliation{National Institute of Chemical Physics and Biophysics, Akadeemia tee 23, 12618 Tallinn, Estonia}

\author{V. Kocsis}
\affiliation{RIKEN Center for Emergent Matter Science (CEMS), Wako 351-0198, Japan}

\author{Y. Tokunaga}
\affiliation{RIKEN Center for Emergent Matter Science (CEMS), Wako 351-0198, Japan}
\affiliation{Department of Advanced Materials Science, University of Tokyo, Kashiwa 277-8561, Japan}

\author{Y. Taguchi}
\affiliation{RIKEN Center for Emergent Matter Science (CEMS), Wako 351-0198, Japan}

\author{Y. Tokura}
\affiliation{RIKEN Center for Emergent Matter Science (CEMS), Wako 351-0198, Japan}
\affiliation{Department of Applied Physics and Tokyo College, University of Tokyo, Tokyo 113-8656, Japan}

\author{I. K{\'e}zsm{\'a}rki}
\affiliation{Department of Physics, Budapest University of Technology and 
	Economics, 1111 Budapest, Hungary}
\affiliation{Experimental Physics V, Center for Electronic Correlations and Magnetism, University of Augsburg, 86135 Augsburg, Germany}

\author{P. Balla}
\affiliation{Institute for Solid State Physics and Optics, Wigner Research Centre or Physics, PO Box. 49, H-1525 Budapest, Hungary}

\author{K. Penc}
\affiliation{Institute for Solid State Physics and Optics, Wigner Research Centre or Physics, PO Box. 49, H-1525 Budapest, Hungary}

\author{J. Romh{\'a}nyi}
\affiliation{Department of Physics and Astronomy, University of California, Irvine, 4129 Frederick Reines Hall, Irvine, CA,USA}

\author{S. Bord{\'a}cs}
\affiliation{Department of Physics, Budapest University of Technology and 
	Economics, 1111 Budapest, Hungary}
\affiliation{Hungarian Academy of Sciences, Premium Postdoctor Program, 1051 
	Budapest, Hungary}

\date{\today}

\begin{abstract}
Non-reciprocal directional dichroism, also called the optical-diode effect, is an appealing functional property inherent to the large class of non-centrosymmetric magnets. However, the in-situ electric control of this phenomenon is challenging as it requires a set of conditions to be fulfilled: Special symmetries of the magnetic ground state, spin-excitations with comparable magnetic- and electric-dipole activity and switchable electric polarization. We demonstrate the isothermal electric switch between domains of \bcgo{} possessing opposite magnetoelectric susceptibilities. Combining THz spectroscopy and multiboson spin-wave analysis, we show that unbalancing the population of antiferromagnetic domains generates the non-reciprocal light absorption of spin excitations.
\end{abstract}

\pacs{Valid PACS appear here}
\maketitle


The interaction between light and matter may produce fascinating phenomena. 
Among them is the non-reciprocal directional dichroism (NDD), when the absorption differs
for the propagation of light along and opposite to a specific direction. 
In contrast to the magnetic circular dichroism, the absorption difference for NDD is finite even for unpolarized light.
The chirality of the light lies at the heart of the phenomenon: the electric ($\mathbf{E}^\omega$) and magnetic ($\mathbf{H}^\omega$) field components of the light and its propagation vector $\mathbf{k} \propto \mathbf{E}^\omega \times \mathbf{H}^\omega$ form a right-handed system. 
Applying orthogonal static electric ($\mathbf{E}$) and magnetic ($\mathbf{H}$) fields to a material breaks the inversion and time-reversal symmetries, leading to the observation of NDD \cite{Rikken2002}. 
Such a symmetry breaking is inherent to magnetoelectric (ME) multiferroics, materials with coexisting electric and magnetic orders. In multiferroics, the ME coupling establishes a connection between responses to electric and magnetic fields: an external electric field generates  magnetization $\mathbf{M}$, and a magnetic field induces electric polarization $\mathbf{P}$ in the sample.
The NDD is manifested by the refractive index difference $\Delta N = N_+-N_-$ for counter-propagating ($\pm\mathbf{k}$) linearly polarized beams~\cite{Kezsmarki2011,Bordacs2012,Kezsmarki2014}.
In the long-wavelength limit
\begin{equation}
N_\pm=\sqrt{\varepsilon_{\alpha\alpha}\mu_{\beta\beta}}\pm\chi_{\alpha\beta}^{em}\;,
\label{eq:refractive_ind}
\end{equation}
where $\varepsilon_{\alpha\alpha}$ and $\mu_{\beta\beta}$ are the components of the permittivity and the permeability tensors for oscillating fields polarized along $E_\alpha^\omega$ \& $H_\beta^\omega$, and $\chi_{\alpha\beta}^{em}$ is the ME susceptibility characterizing the induced polarization $\delta P_\alpha^\omega \propto \chi_{\alpha\beta}^{em}H_\beta^\omega$. The $\chi_{\alpha\beta}^{em}$ becomes resonantly enhanced for spin excitations of multiferroics  endowed with a mixed magnetic and electric dipole character giving rise to strong NDD \cite{Kezsmarki2011,Bordacs2012,Takahashi2012,Kezsmarki2014,Szaller2014SumRule,Kibayashi2014MChDhelimagnet,Kuzmenko2015NDDferroborate,Yu2018FeZnMo3O8NDD}.


Since $\Delta N \propto \chi_{\alpha\beta}^{em}$, the absorbing and transparent directions are determined by the sign of ME susceptibility, and therefore, they can be interchanged by the sign reversal of the $\chi_{\alpha\beta}^{em}$. 
The magnetic field can naturally switch between time-reversed magnetic states with opposite signs of ME responses, and allows the control of NDD \cite{Kezsmarki2011,Bordacs2012,Kezsmarki2014}.
Can we achieve a similar switch with an electric field, which is a time-reversal even quantity? 
Apart from being a fundamental question, the voltage control of NDD may promote the application of multiferroics in GHz-THz frequency data transmission and signal processing devices with reduced size and energy consumption. In addition to the NDD, the electric field induced switching between time-reversed magnetic states would also provide an efficient way to control other optical ME effects, such as chirality of magnons \cite{Bordacs2012,Kuzmenko2014OpticalActivity,Masuda2021Econtrol} or axion-term-induced gyrotropy \cite{Kurumaji2017Axion}. 
The ME coupling may help us to achieve the desired control of magnetic states \cite{Krichevtsov1986Nonrecpolrot,Saito2009EfieldRot,Sato2020MChD,Takahashi2012,Kibayashi2014MChDhelimagnet,Kocsis2018LiCoPO4}, however realizing this effect is not at all trivial. It requires a magnetic order permitting NDD and a polarization that is switchable by laboratory electric fields. In the visible spectral range, the realization of this effect has been confirmed for charge excitations \cite{Saito2009EfieldRot,Sato2020MChD}. However, studies in the THz range of spin-wave excitations are scarce. So far, mostly ME poling was used to select between time-reversed domains by cooling the sample through the ordering temperature in external magnetic and electric fields~\cite{Takahashi2012,Kibayashi2014MChDhelimagnet,Kocsis2018LiCoPO4}. The electric field induced changes in the absorption coefficient was detected only recently \cite{Kuzmenko2018BoratesPRL}.
 

In this letter, we demonstrate the isothermal electric field control of the THz frequency NDD in \bcgo{}, which provides an ideal model system due to its simple antiferromagnetic (AFM) order.
The electric field switches between the transparent and absorbing directions, where the absorption difference between the two is found experimentally as high as 30\%.  
We attribute the observed change of the NDD to the electric field induced imbalance in the population of the AFM domains.

\begin{figure}
	\centering
	\includegraphics[width=\columnwidth]{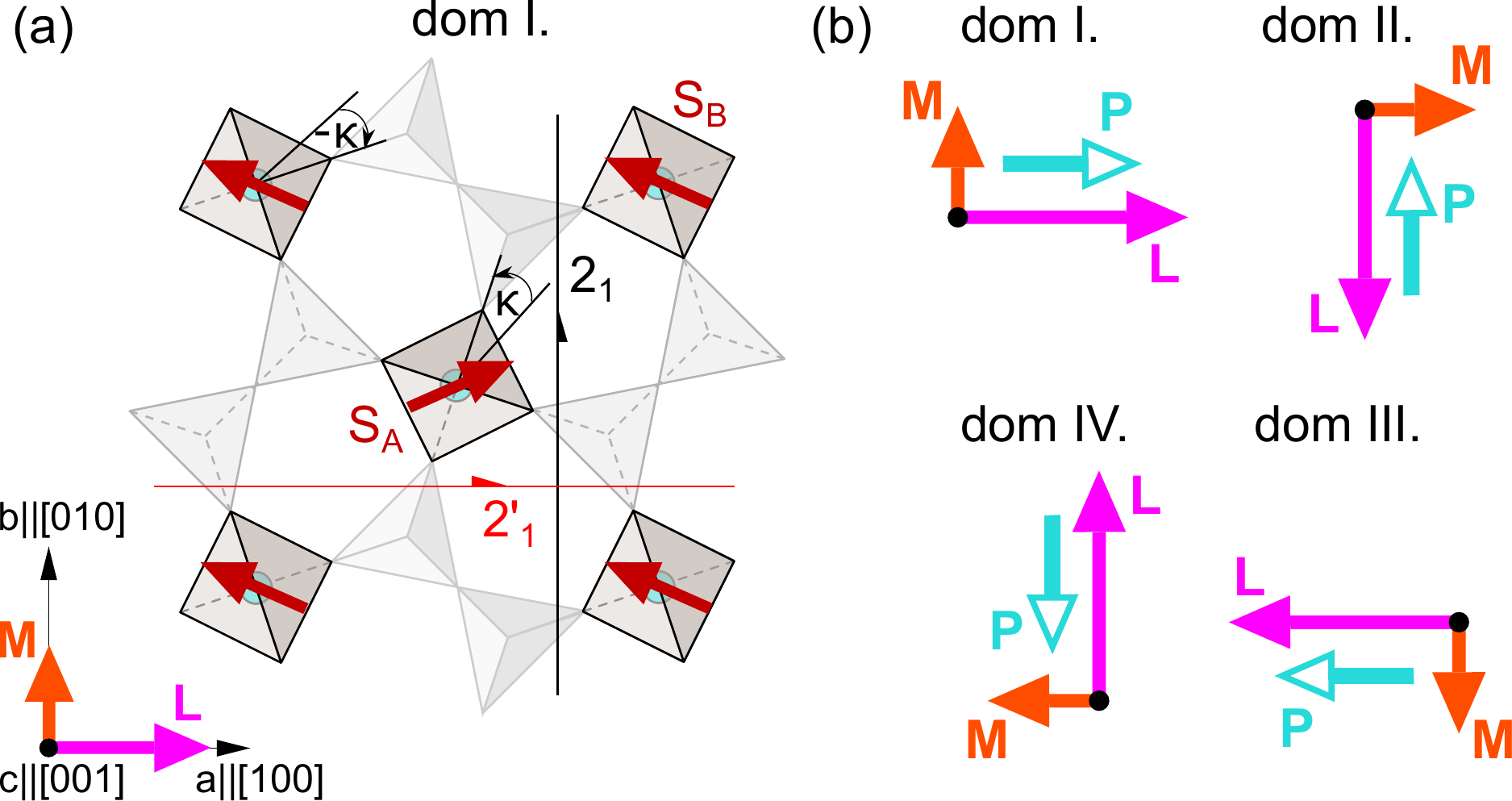}
    \caption{(a) The canted antiferromagnetic order of \bcgo{} in domain I in zero fields. Cyan circles denote the Co$^{2+}$ ions with $S=3/2$ (dark red arrows) in the center of the O$^{2-}$ tetrahedra (grey). The symmetry operations are the $2_1$ screw axis, (black  half-arrow) and the orthogonal $2_1'$ screw axis followed by the time-reversal (red half-arrow). $\mathbf{M}$ and $\mathbf{L}$ correspond to the uniform and staggered sublattice magnetizations, respectively. (b) The four antiferromagnetic domains. A magnetic field applied along the $[001]$ axis induces a polarization $\delta\mathbf{P}$ (light blue arrows) via linear magnetoelectric effect.}
	\label{fig:BCGO_domains}
\end{figure}


The discovery of the ME properties of \bcgo{}~\cite{Yi2008}, followed by a detection of the gigantic ME effect in Ca$_{2}$CoSi$_{2}$O$_{7}$ \cite{Akaki2009} aroused interest in this family of quasi-two dimensional compounds. 
They crystallize in the non-centrosymmetric $P\overline{4}2_1m$ structure, where the unit cell includes two spin-3/2 magnetic Co$^{2+}$ ions, as shown in Fig.~\ref{fig:BCGO_domains}(a). 
Below  $T_N$=6.7\,K, the spins order in a two-sublattice easy-plane AFM structure \cite{Zheludev2003PRB_INS}. 
A small in-plane anisotropy pins the AFM ordering vector ($\mathbf{L}=\mathbf{M}_A-\mathbf{M}_B$) to one of the symmetry-equivalent $\langle$100$\rangle$ directions of the tetragonal plane, as shown in Fig.~\ref{fig:BCGO_domains} \cite{Romhany2011AFEPolarization,Soda2014PRL,Soda2016PolND}.  
Applying an external magnetic field $\mathbf{H}\!\!\parallel\!\![110]$ rotates the $\mathbf{L}$ vector to $[1\overline{1}0]$, and gives rise to a sizeable ferroelectric polarization $\mathbf{P}$ along the tetragonal [001] axis~\cite{Murakawa2010PRL}. 
The same ME interaction leads to NDD for the THz spin excitations of \bcgo{} \cite{Miyahara2011JPSJ,Penc2012BCGOPRL}, which has been observed for in-plane magnetic fields: i) for light propagation $\mathbf{k}$ along the cross product of the magnetic field $\mathbf{H}\parallel$[110] and the magnetic-field-induced polarization $
\mathbf{P}\parallel$[001] \cite{Kezsmarki2011,Kezsmarki2014}, and ii) for $\mathbf{k}\parallel\mathbf{H}\parallel$[100] when a chiral state is realized~\cite{Bordacs2012}.

Both the static and the dynamic ME response of \bcgo{} are consistently explained by the spin-dependent $p$-$d$ hybridization \cite{Arima2007,Murakawa2010PRL,Yamauchi2011DFT,Miyahara2011JPSJ,Penc2012BCGOPRL}. In this mechanism, the spin-quadrupole operators of the $S=3/2$ cobalt spin directly couple to the induced polarization $\mathbf{P}_j$,
 \begin{align}
 	P^{a}_{j} &\propto
 	-\cos2\kappa_j\left( S^b_{j} S^c_{j} \!+\! S^c_{j} S^b_{j}\right)
 	+\sin2\kappa_j\left( S^a_{j} S^c_{j} \!+\! S^c_{j} S^a_{j}\right),
 	\nonumber\\
 	P^{b}_{j}&\propto
 	-\cos2\kappa_j\left( S^a_{j} S^c_{j} \!+\! S^c_{j} S^a_{j}\right)
 	-\sin2\kappa_j\left( S^b_{j} S^c_{j} \!+\! S^c_{j} S^b_{j}\right),
 	\nonumber\\
 	P^{c}_{j}&\propto
 	-\cos2\kappa_j\left( S^a_{j} S^b_{j} \!+\! S^b_{j} S^a_{j}\right)
 	+\sin2\kappa_j\left(( S^a_{j})^2 \!-\! ( S^b_{j})^2\right),
 	\label{eq:pol}
 \end{align}
where $j$ is the site index, and $a$, $b$, $c$ are parallel to [100], [010] and [001], respectively. $\kappa_j=\kappa$ in A sublattice and $\kappa_j=-\kappa$ in B sublattice account for the different orientation of the tetrahedra [see Fig.~\ref{fig:BCGO_domains}(a)]. 
The same mechanism is the source of the multiferroic properties of Sr${}_{2}$CoSi${}_{2}$O${}_{7}$ \cite{Akaki2012}, the observation of spin-quadrupolar excitations in Sr${}_{2}$CoGe${}_{2}$O${}_{7}$ in the field aligned phase \cite{Akaki2017}, and the microwave nonreciprocity of magnons in Ba$_2$MnGe$_2$O$_7$ \cite{Iguchi2018}.

The clue how to control the NDD using electric fields comes from the experiment of Murakawa et al.~\cite{Murakawa2010PRL}. They showed that a magnetic field applied nearly parallel to the tetragonal axis induces an in-plane electric polarization along one of the $\langle$100$\rangle$ directions. 
The hysteresis of the polarization observed upon tilting the field away from the [001] axis suggests a rearrangement of the magnetic domain population.
The AFM order reduces the space group symmetry from $P\overline{4}2_1m1'$ to $P2_1'2_12'$, corresponding to the breaking of the rotoreflection symmetry $\overline{4}$, and the formation of four magnetic domains, shown in Fig.~\ref{fig:BCGO_domains} (b) \footnote{The number of the domains is determined by the order of the factor group $S_4 \cong P\overline{4}2_1m1'/P2_1'2_12'$, which also transforms the domain states among each other.}.
The $P2_1'2_12'$ symmetry gives rise to a finite $\chi^{em}$ and in a magnetic field $\mathbf{H}\!\parallel\! [001]$ a  polarization $\delta\mathbf{P}$ parallel to the $\mathbf{L}$ develops, as shown in Fig.~\ref{fig:BCGO_domains}(b).
If the field is perfectly aligned $\mathbf{H}\parallel$[001], the four domains remain equivalent and the field-induced polarizations $\delta\mathbf{P}$ cancel out.
However, a small perturbation such as tilting of the magnetic field or applying an in-plane electric field can break the delicate balance between the domains. In our experiments, we exploit this highly susceptible state to change the relative population of the domains by electric field, $\mathbf{E}\!\parallel\![100]$, and attain control over the NDD, present for $\mathbf{E}\times\mathbf{H}$.


\begin{figure}[h!]
	\centering
	\includegraphics[width=\columnwidth]{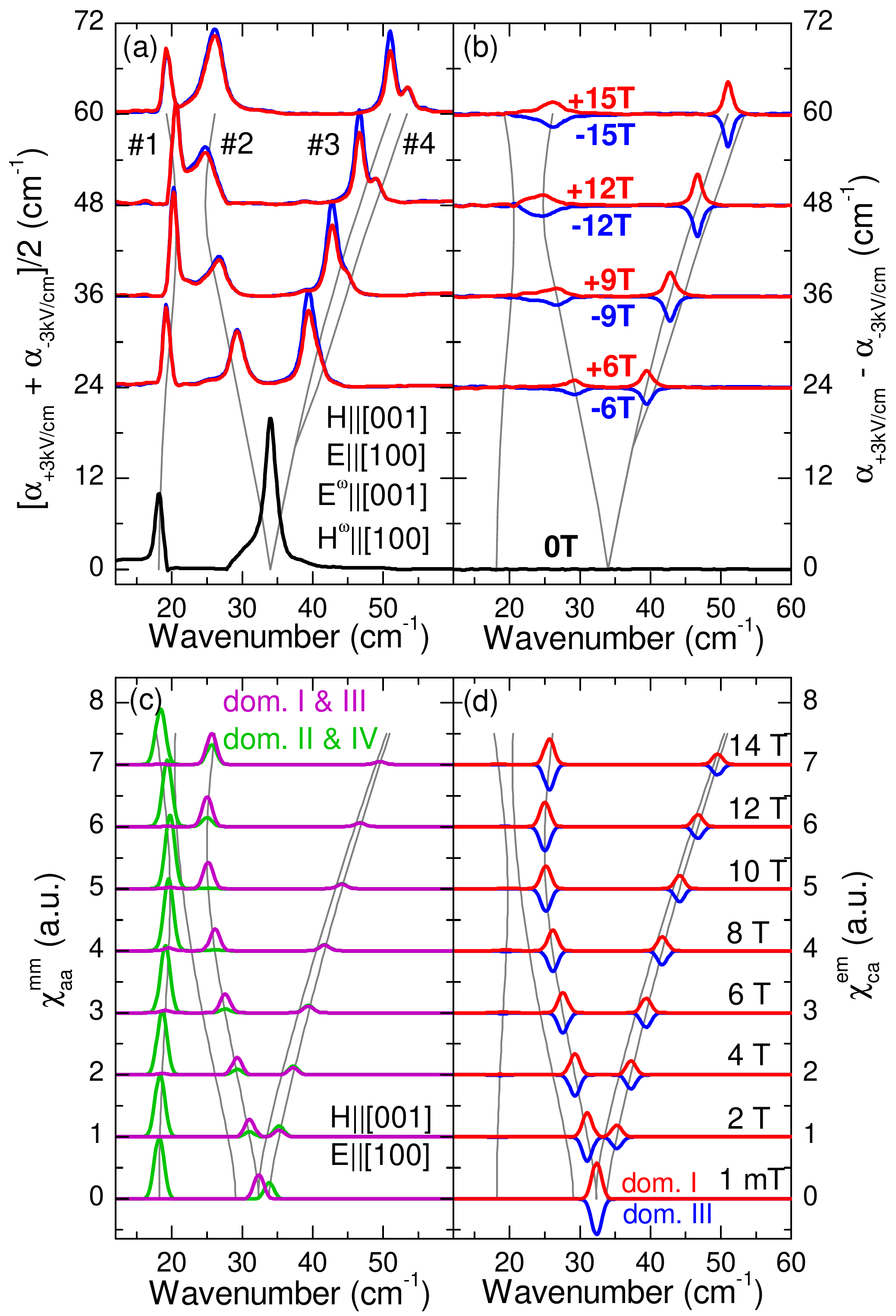}
    \caption{(a) Magnetic field dependence of the THz absorption spectra averaged for the measurements performed in electric fields with opposite signs, $E$=$\pm$3\,kV/cm at $T$=3.5\,K. The light polarization is $\mathbf{E}^\omega\parallel$[001] and $\mathbf{H}^\omega\parallel$[100]. The spectra measured in positive(red)/negative(blue) magnetic fields $\mathbf{H}\parallel$[001] are shifted in proportion with the absolute value of the field. Grey lines indicate the magnetic field dependence of the resonance energies. (b) shows the electric field-induced change in the absorption spectra as the difference of the absorption spectra recorded in $E$=$\pm$3\,kV/cm. (c) The magnetic susceptibility calculated from the spin-wave theory in domain I \& III (purple) and in domain II \& IV (green). (d) The ME susceptibility in domain I (red) and domain III (blue).}
	\label{fig:spectra_Bdep}
\end{figure}

 \bcgo{} single crystals were grown by the floating zone technique as described in \cite{Murakawa2010PRL}. Silver paste electrodes were painted on the parallel sides of a 2x3x0.7\,mm$^3$ rectangular (010) cut. The THz spectra were measured in Tallinn with a Martin-Puplett interferometer and a 0.3\,K silicon bolometer. We applied the external magnetic and electric fields in the $\mathbf{H}\parallel$[001] and $\mathbf{E}\parallel$[100] directions, while the THz radiation propagated along the $\mathbf{k}\parallel$[010] axis. The crystallographic axes of the sample were oriented by X-ray Laue diffraction and aligned in the THz experiment by at least to 1$^\circ$ precision. The THz absorption spectra were deduced as described in Ref.~\cite{Kezsmarki2015}.

Our main experimental results are summarized in Fig.~\ref{fig:spectra_Bdep}. Panel (a) displays the average and (b) the difference of the THz absorption spectra measured in electric fields with opposite signs ($E$=$\pm$3\,kV/cm) and constant magnetic fields. 
In agreement with former results \cite{Penc2012BCGOPRL}, we assign the absorption peak around 18\,cm$^{-1}$ (mode \#1) to the optical magnon excitation of the easy-plane AFM ground state whereas resonances \#2, \#3 and \#4, showing a V-shape splitting in magnetic fields, are attributed to the spin stretching modes involving the modulation of the spin length. In a finite magnetic field, the absorption spectra become different for the opposite signs of the electric field as evidenced by Fig.~\ref{fig:spectra_Bdep}(b) for the light polarization $\mathbf{E}^\omega\!\parallel\![001]$ and $\mathbf{H}^\omega\!\parallel\![100]$. The electric field odd component of the signal is the manifestation of the NDD and it shows that the absorption is different for light propagation along or opposite to the cross-product of the static electric and magnetic fields $\mathbf{E}\times\mathbf{H}$. This relation is further supported by the fact that the differential absorption spectra change sign under the reversal of the external magnetic field. The NDD is finite only for the spin stretching modes \#2 and \#3 and it increases with magnetic fields up to $\sim$12\,T. We note that for the orthogonal light polarization, $\mathbf{E}^\omega\parallel$[100] and $\mathbf{H}^\omega\parallel$[001], we did not find electric field induced absorption difference within the accuracy of the experiment.

\begin{figure}
	\centering
	\includegraphics[width=\columnwidth]{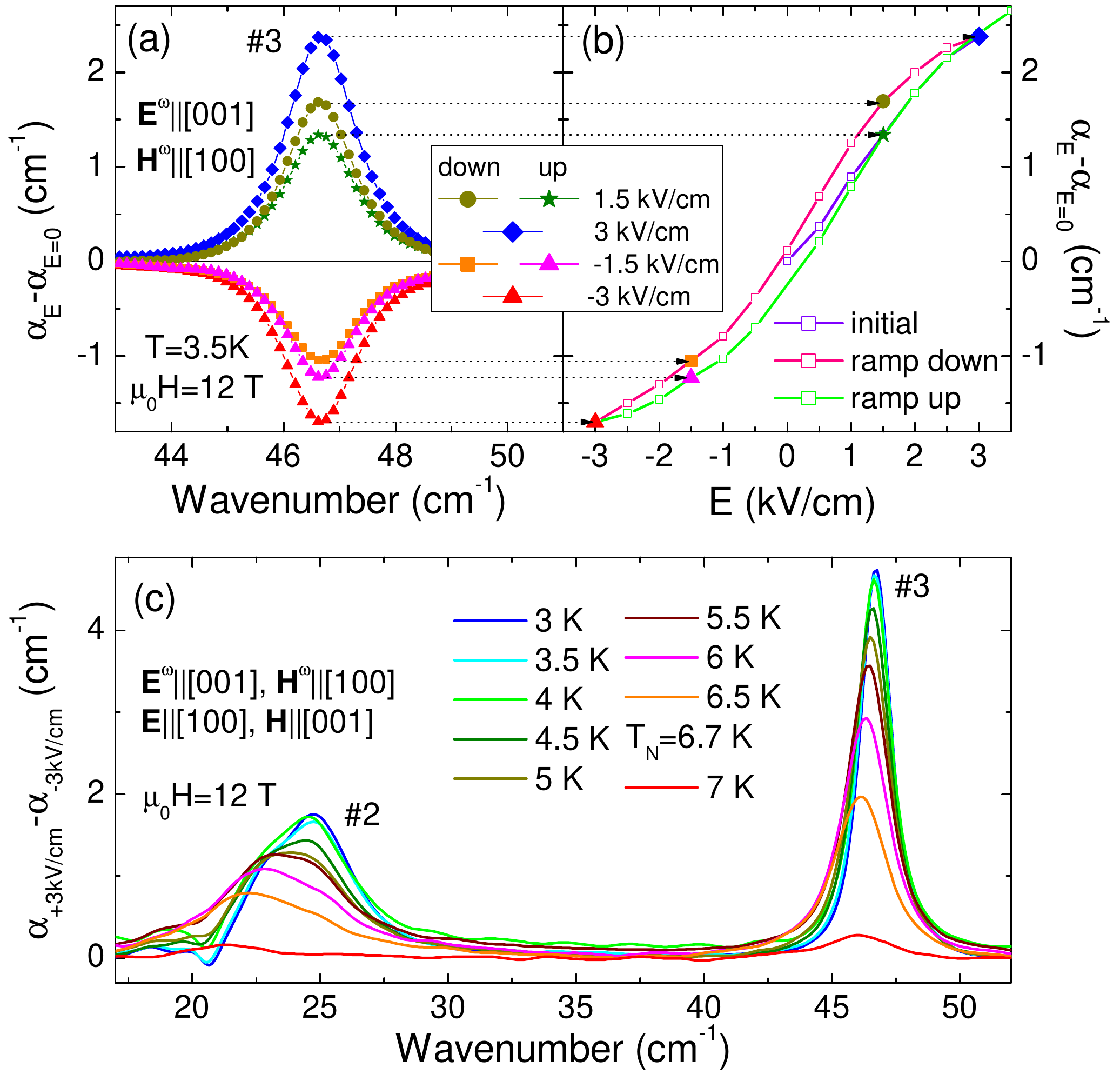}
    \caption{(a) The electric field induced change in the absorption spectra measured with respect to the zero field cooled state at 3.5\,K and in fixed magnetic field 12\,T. (b) The hysteresis of the electric field dependence of the peak absorption. The horizontal arrows connect corresponding points of panels (a) and (b). (c) Temperature dependence of the electric field induced change in the absorption spectra measured in 12\,T.}
	\label{fig:spectra_Edep_Tdep}
\end{figure}

The electric field induced change in the absorption spectra around mode \#3, measured with respect to the zero field cooled state, is displayed in Fig.~\ref{fig:spectra_Edep_Tdep}(a). The peak absorption, shown in Fig.~\ref{fig:spectra_Edep_Tdep}(b), depends on the electric field history of the sample: the initial and the following upward and downward sweeps are all different and the absorption difference has a small but finite remanence \cite{Supplement}. Furthermore, the electric field can change the absorption only below $T_N$ as displayed in Fig.~\ref{fig:spectra_Edep_Tdep}(c), though the intensity of the spin stretching mode remains finite even above $T_N$ \cite{Kezsmarki2011}. All of these findings suggest that the observed electric field effect arises only in the magnetically ordered phase and it is related to switching between domain states possessing different NDD.  

Considering the symmetries of the zero-field ground state shown in Fig.~\ref{fig:BCGO_domains}(a), the (unitary) $2_1$ screw axis restricts NDD for light propagation 
$\mathbf{k}\!\parallel\! \mathbf{c}\times\mathbf{L}$ 
in a given domain. When a magnetic field is applied along $\mathbf{H}\!\parallel\![001]$, only the $2_1'$ symmetry remains. 
The $S^b$, $S^c$, $P^a$ operators are even, while $S^a$, $P^b$, $P^c$ are odd under $2_1'$ in domain I, depicted in Fig.~\ref{fig:BCGO_domains}(a).
 Since time reversal makes this symmetry antiunitary, the operators are either even or odd under conjugation, restricting the transition matrix elements to be either real or imaginary~\cite{Viirok2019,Supplement}. 
 As a consequence, the real part of a ME susceptibility combined  from an even and odd operator vanishes, annulling the time-reversal odd part of $\chi^{em}_{bc}$ and $\chi^{em}_{cb}$, thus forbidding NDD when $\mathbf{k}\!\parallel\!\mathbf{H}\!\parallel\![100]$.
  The $2_1'$ does not affect NDD in the other propagation directions, and indeed this is what we observed for $\mathbf{k}\!\parallel\![010]$. In finite fields, we also expect NDD for the $\mathbf{k}\!\parallel\![001]$ -- but then the analysis of results would be more complicated as the Faraday effect mixes the polarization states of the light. 


In order to interpret the experimental results quantitatively, we considered the microscopic Hamiltonian of interacting S=3/2 Co$^{2+}$ spins following Refs.~\cite{Miyahara2011JPSJ,Penc2012BCGOPRL}:
\begin{align}
\mathcal{H}=&\sum\limits_{\langle i,j\rangle}[J(\hat{S}_i^a\hat{S}_j^a+\hat{S}_i^b\hat{S}_j^b)+J^c\hat{S}_i^c\hat{S}_j^c]+\sum\limits_{i}\Lambda(\hat{S}_i^c)^2 \nonumber \\
 & -\sum\limits_{i}[g_{cc}H_c\hat{S}_i^c+E_a\hat{P}_i^a], 
 \label{eq:Hamilton}
\end{align}
where summation $\langle i,j\rangle$ runs over the nearest neighbours. Beside the anisotropic exchange coupling ($J$ and $J^c$), single-ion anisotropy $\Lambda$, and the Zeeman term, we introduce the coupling between the external electric field, $E_a$, and the spin-induced polarization (see Eq.~\ref{eq:pol}), which breaks the O(2) symmetry of the model.

We calculated the excitations above a variational site-factorized ground state using a multiboson spin-wave theory, following Ref.~\onlinecite{Penc2012BCGOPRL}. The approximate O(2) symmetry of the Hamiltonian (even for finite $H_c$) is reflected in the ground state manifold, the application of a tiny $E_a>0$ combined with $H_c>0$ selects domain I in Fig.~\ref{fig:BCGO_domains}(b), while $E_a<0$ selects domain III as the variational ground state. We note that even in the highest fields the ME energy \cite{Murakawa2010PRL} is at least an order of magnitude smaller than the in-plane anisotropy \cite{Soda2014PRL}, thus, the rotation of the AFM vector $\mathbf{L}$ away from the principal axes is negligible (e.g.~\cite{Soda2016PolND}). The magnetic dipole strengths of the excitations are estimated by the transition matrix elements of the spin operators $|\langle 0|\hat{S}^\alpha|n \rangle|^2$ between the ground state $|0 \rangle$ and the excited states $|n \rangle$. The contribution of the magnetic dipole processes to the absorption is shown in Fig.~\ref{fig:spectra_Bdep}(c). The electric dipole matrix elements are evaluated similarly for polarization components $\hat{P}^\beta$. The ME susceptibility, $\chi_{ca}^{em}\propto\langle 0|P_c|n \rangle\langle n|S_a|0 \rangle$ is plotted in Fig.~\ref{fig:spectra_Bdep}(d). 

For light polarization $\mathbf{E}^\omega\!\!\parallel\!\![001]$ and $\mathbf{H}^\omega\!\parallel\![100]$, our model predicts that two spin stretching modes have finite ME susceptibility $\chi_{ca}^{em}$ and correspondingly show NDD with the same sign. The overall sign of the ME response is reversed upon the reversal of either the static electric or the magnetic field related to the switching from domain I to III. All of these findings are in agreement with the experiments and imply that the electric field control of the NDD is realized by influencing the AFM domains. 
We note that among modes \#3 and \#4, which show a tiny splitting in high fields, resonance \#3 is NDD active in the experiment, whereas our theory predicts NDD for the higher energy mode. However, we found no obvious way to reproduce the fine structure of the resonance energies within our model or by including other realistic terms \cite{Murakawa2010PRL,Miyahara2011JPSJ,Romhany2011AFEPolarization}. 

Although theory predicts that individual domains possess a finite dichroism as $H_c\to 0$ [see Fig.~\ref{fig:spectra_Edep_Tdep}(b)], we observed vanishing NDD in this limit. This suggests that domain walls relax toward their initial positions and the domain population evens out as fields go to zero. The multidomain state may be favored by: (i) electric dipole-dipole interaction between the ferroelectric domains; (ii) elastic energy, since the AFM domains break the tetragonal symmetry they can couple to orthorhombic distortion \cite{Nakajima2015Stress}.
The finite intensity of mode \#1 also indicates that domains II \& IV coexist with domain I \& III. In domain I \& III, excitation \#1 is silent for this light polarization according to the calculation, since it can only be excited by the $\mathbf{H}^\omega\!\parallel\![010]$, which is perpendicular to $\mathbf{L}\!\parallel\![100]$. The polarization matrix element is also negligible for this resonance. Therefore, domains II \& IV with finite magnetic dipole strength for $\mathbf{H}^\omega\!\parallel\![100]$[see Fig.~\ref{fig:spectra_Bdep}(c)] should also be present in the studied sample. 
Thus, one expects even stronger NDD than observed experimentally here, if the mono-domain state of either domain I or domain III can be realized.
Finally, we note that the small difference in the averaged absorption [Fig.~\ref{fig:spectra_Bdep}(a)] observed for the reversal of the magnetic field is probably caused by a small misalignment. When the magnetic field is slightly tilted toward the light propagation $\mathbf{k}\parallel$[010], the balance between domain I and III can be broken.

The absence of the NDD for the orthogonal light polarization, $\mathbf{E}^\omega\!\parallel\![100]$ and $\mathbf{H}^\omega\!\parallel\![001]$, can be explained by the smallness of the $\chi^{em}_{ac}$. Due to the nearly preserved O(2) symmetry of the system, the magnetic matrix element in $\chi^{em}_{ac}$ involves the $\hat{S}^c$, which commutes with the terms of the Hamiltonian in Eq.~\ref{eq:Hamilton} except for the $\mathbf{E}\cdot\mathbf{P}$. Therefore the dipole oscillator strength for $S^c$ -- given by the double commutator \cite{Hohenberg1974} -- is tiny compared to other matrix elements.


In summary, we demonstrated the isothermal voltage control of the non-reciprocal THz absorption in \bcgo{}. In contrast to former studies applied ME poling, here the ME polarization is induced by a magnetic field preserving the nearly degenerate ground states within the tetragonal plane. This manifold allows efficient voltage control of the magnetic domain population and so of the NDD. A similar mechanism may give rise to NDD in ME spin-spiral compounds e.g.~Cu$_2$OSeO$_3$ or CoCr$_2$O$_4$ with multi-domain states. Our results can promote the applications of multiferroics in voltage-controlled high-frequency devices and stimulate search for compounds with stronger remanence and higher ordering temperatures.

{\it Note added}: During the preparation of this manuscript, we become aware of the related work of Kimura et al., who study the electric field control of microwave NDD of the triplet Bose-condensate in TlCuCl$_3$ \cite{Kimura2020}.

\begin{acknowledgments}
The authors thank M. Mostovoy for enlightening discussions. This research was supported by the Estonian Ministry of Education and Research grants IUT23-3, PRG736, by the European Regional Development Fund Project No. TK134, by the bilateral program of the Estonian and Hungarian Academies of Sciences under the Contract NMK2018-47, by the Hungarian National Research, Development and Innovation Office – NKFIH grants ANN 122879 and K 124176, and by the Hungarian ELKH. The research reported in this paper and carried out at the BME has been supported by the NRDI Fund (TKP2020 IES, Grant No. BME-IE-NAT) based on the charter of bolster issued by the NRDI Office under the auspices of the Ministry for Innovation and Technology. J. V\'it was partially supported by the Grant Agency of the Czech Technical University in Prague (Project No. SGS19/188/OHK4/3T/14) and by the project SOLID21 (Project No. CZ.02.1.01/0.0/0.0/16 019/0000760).
\end{acknowledgments}


%


\onecolumngrid
\hspace{8 cm}
\begin{center}
\textbf{\large Supplementary material: \textit{In Situ} Electric Field Control of THz Nonreciprocal Directional Dichroism in the Multiferroic \bcgo{}}
\end{center}
\twocolumngrid
\setcounter{equation}{0}
\setcounter{figure}{0}
\setcounter{table}{0}
\setcounter{page}{1}
\makeatletter
\renewcommand{\theequation}{S\arabic{equation}}
\renewcommand{\thefigure}{S\arabic{figure}}
\renewcommand{\bibnumfmt}[1]{[S#1]}
\renewcommand{\citenumfont}[1]{S#1}

\section{I. Magnetoelectric annealing}

To study the effect of magnetoelectric (ME) annealing (poling), we cooled the sample from 10\,K to 3.5\,K in the presence of electric, $E_{poling}$ and magnetic fields $H_{poling}$. At low temperature we measured the THz absorption in finite electric, $E$ and magnetic fields $H$. The magnitude of the electric and magnetic fields applied for annealing or during the measurements were 3\,kV/cm and 8\,T, respectively. When we turned off the fields the domain population became nearly uniform as discussed in the main text. 

\begin{figure}
	\centering
	\includegraphics[width=\columnwidth]{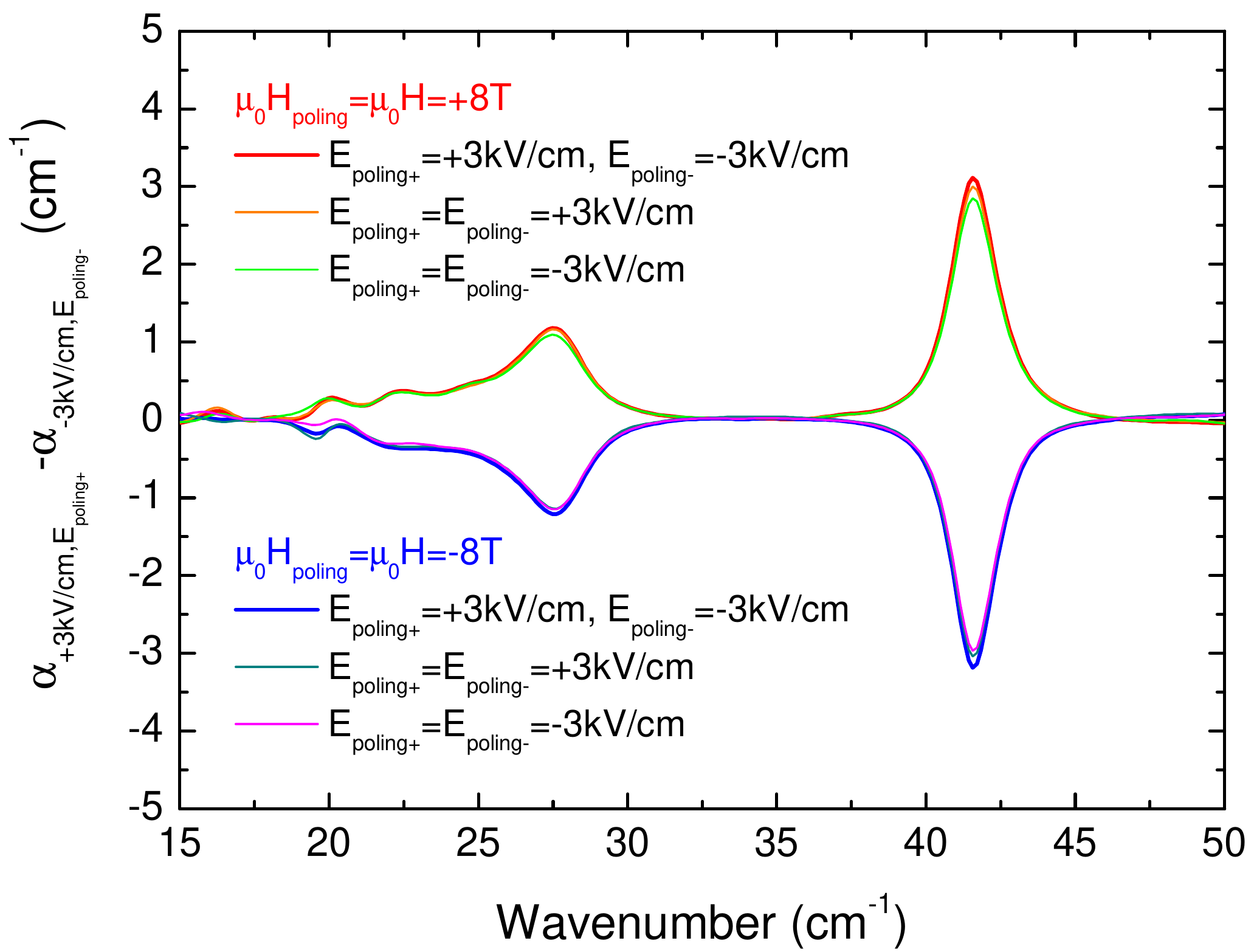}
    \caption{Magnetoelectric annealing dependence of the electric field-induced change in the THz absorption spectra measured at $T$=3.5\,K in $\mathbf{H}\parallel$[001]. The light polarization is $\mathbf{E}^\omega\parallel$[001] and $\mathbf{H}^\omega\parallel$[100]. $E_{poling+}$ and $E_{poling-}$ are the electric fields applied during the annealing procedure before measuring the absorption in positive and negative electric fields, respectively. For detailed description of the field history see the text.} 
	\label{fig:poling}
\end{figure}

The difference of the absorption spectra measured in $E$=$E_{poling}$ with opposite signs at a fixed $H_{poling}$ right after cooling to 3.5\,K are shown in red and blue in Fig.~\ref{fig:poling}. After the annealing we reversed the electric field at low temperature and measured the absorption difference again. Orange and green curves are recorded for $\mu_0H_{poling}$=+8\,T and positive and negative poling electric fields, respectively. Dark cyan and magenta curves are measured for $\mu_0H_{poling}$=-8\,T and positive and negative poling electric fields, respectively. The absorption difference is almost identical for ME annealing and in-situ reversal of the electric fields. These measurements confirm that at low temperature we can control the antiferromagnetic domains as much as the ME term allows. 

\section{II. Selection rules of the magnetoelectric susceptibility}

 In this section we provide details for the symmetry classification of magnetization and electric polarization operators that determine the NDD response in the ordered state. The classification uses the symmetries of the antiferromagnetically ordered state in Ba$_2$CoGe$_2$O$_7$ first in zero, then in finite magnetic field applied along the $[001]$ direction, corresponding to the experiments. 
 
 Here we choose the directions of the coordinate axes in spin and polarization space according to the crystallographic ones, i.e. $a||\left[100\right]$,  $b||\left[010\right]$,  and $c||\left[001\right]$.  
\setlength{\extrarowheight}{2pt}
\begin{table}[b]
	\caption{\label{table:transformation properties} Transformation properties of the site indices, polarization and spin (magnetization) components. The components $(P^a,P^b,P^c)$ transform the same way as the coordinates $(x,y,z)$, and the components of the magnetization $(M^a,M^b,M^c)$ transform the same way as the spin components $(S^a_j,S^b_j,S^c_j)$.}
	\begin{ruledtabular}
		\begin{tabular}{ c c c c }
			$\mathbb{1}$	& $\tilde{C}_2^{b}$ & $\Theta \tilde{C}_2^{a}$ & $\Theta C_2^{c}$ \\ \hline
			$A$	& $B$ & $B$ & $A$ \\
			$B$	& $A$ & $A$ & $B$ \\ \hline
			$P^a$ & $-P^a$ & $+P^a$ & $-P^a$ \\
			$P^b$ & $+P^b$ & $-P^b$ & $-P^b$ \\
			$P^c$ & $-P^c$ & $-P^c$ & $+P^c$ \\ \hline
	        $\Pi^a$ & $+\Pi^a$ & $-\Pi^a$ & $-\Pi^a$ \\
			$\Pi^b$ & $-\Pi^b$ & $+\Pi^b$ & $-\Pi^b$ \\
			$\Pi^c$ & $+\Pi^c$ & $+\Pi^c$ & $+\Pi^c$ \\ \hline
	        $M^a$ & $-M^a$ & $-M^a$ & $+M^a$ \\
			$M^b$ & $+M^b$ & $+M^b$ & $+M^b$ \\
			$M^c$ & $-M^c$ & $+M^c$ & $-M^c$ \\ \hline
			$L^a$ & $+L^a$ & $+L^a$ & $+L^a$ \\
			$L^b$ & $-L^b$ & $-L^b$ & $+L^b$ \\
			$L^c$ & $+L^c$ & $-L^c$ & $-L^c$ 
		\end{tabular}
	\end{ruledtabular}
\end{table}

\subsection{Zero external magnetic field}

Without external magnetic field the generators of the magnetic space group in the domain I [shown in Fig. 1(a) in the main text] are isomorphic to the magnetic point group  
\begin{align}
	\left\{\mathbb{1}, \{C_2^b|[0\frac{1}{2}0]\} ,\{\Theta C_2^{a}|[\frac{1}{2}00]\} ,  \{\Theta C_2^{c}|[000]\}\right\}
	\cong\nonumber\\ 
	\cong\left\{\mathbb{1}, \tilde{C}_2^{b}, \Theta \tilde{C}_2^{a}, \Theta C_2^{c} \right\}.
	\label{eq:D2C2}
\end{align}
This magnetic (non-unitary) point group is called $22'2'$ or $D_2(C_2)$.  The operators with a tilde exchange sites $A$ and $B$ and $\Theta$ denotes the time reversal operation.

\setlength{\extrarowheight}{2pt}
\begin{table}[t]
	\caption{\label{table:B_zero_irreps} Character table and transformation properties of the uniform and staggered magnetization and polarization components in the absence of external fields. $A_1$ is the identity irreducible representation.}
	\begin{ruledtabular}
		\begin{tabular}{ c rrrr c}		
			Irrep	& $\mathbb{1}$	& $\tilde{C}_2^{b}$ & $\Theta \tilde{C}_2^{a}$ & $\Theta C_2^{c}$ & Operators\\ \hline
			$A_1$	& $1$ & $1$ & $1$ & $1$ & $M^b$, $L^a$, $\Pi^c$ \\
			$A_2$	& $1$ & $1$ & $-1$ & $-1$ & $L^c$, $P^b$, $\Pi^a$ \\  
		    $B_1$	& $1$ & $-1$ & $1$ & $-1$ & $M^c$, $P^a$, $\Pi^b$ \\
			$B_2$	& $1$ & $-1$ & $-1$ & $1$ & $M^a$, $L^b$, $P^c$ \\  
		\end{tabular}
	\end{ruledtabular}
\end{table}

\setlength{\extrarowheight}{2pt}
\begin{table}[b]
	\caption{\label{table:B_zero_matrix_elements} Matrix elements in zero external magnetic field, based on the character table of $D_2(C_2)$, Table.~\ref{table:B_zero_irreps}. Since the magnetic and electric components of the light are perpendicular, the cases when $\mathbf{H}^{\omega} \| \mathbf{E}^{\omega}$ are meaningless and are denoted by crosses ($\times$). }
	\begin{ruledtabular}
		\begin{tabular}{ cccccc}		
&& $\mathbf{H}^{\omega}\|[100]$ & $\mathbf{H}^{\omega}\|[010]$ & $\mathbf{H}^{\omega}\|[001]$ \\
 &irreps.& $B_2$ & $A_1$ & $B_1$ \\
\hline
$\mathbf{E}^{\omega}\|[100]$ & $B_1$ & $\times$ & 0 & finite (small) \\
$\mathbf{E}^{\omega}\|[010]$ & $A_2$ & 0 & $\times$ & 0\\
$\mathbf{E}^{\omega}\|[001]$ &  $B_2$ & finite & 0 & $\times$ \\
		\end{tabular}
	\end{ruledtabular}
\end{table}

  Let us denote the uniform components of the magnetization and polarization by 
 \begin{subequations}
	\begin{align}
		\mathbf{M} &= \mathbf{S}_A+\mathbf{S}_B, \\
		\mathbf{P} &= \mathbf{P}_A+\mathbf{P}_B,\label{eq:magn_pol_1_uni}
	\end{align}
\end{subequations}  
and the staggered components by
\begin{subequations}
	\begin{align}
		\mathbf{L} &= \mathbf{S}_A-\mathbf{S}_B, \\
        \mathbf{\Pi} &= \mathbf{P}_A-\mathbf{P}_B,
	\end{align}
\end{subequations}
 where $\mathbf{S}_A$ is the spin on site $A$, $\mathbf{P}_A$ is the polarization on site $A$, and so on.  
The polarization components are 
 \begin{align}
 	P^{a}_{j} &\propto
 	-\cos2\kappa_j\left( S^b_{j} S^c_{j} \!+\! S^c_{j} S^b_{j}\right)
 	+\sin2\kappa_j\left( S^a_{j} S^c_{j} \!+\! S^c_{j} S^a_{j}\right),
 	\nonumber\\
 	P^{b}_{j}&\propto
 	-\cos2\kappa_j\left( S^a_{j} S^c_{j} \!+\! S^c_{j} S^a_{j}\right)
 	-\sin2\kappa_j\left( S^b_{j} S^c_{j} \!+\! S^c_{j} S^b_{j}\right),
 	\nonumber\\
 	P^{c}_{j}&\propto
 	-\cos2\kappa_j\left( S^a_{j} S^b_{j} \!+\! S^b_{j} S^a_{j}\right)
 	+\sin2\kappa_j\left(( S^a_{j})^2 \!-\! ( S^b_{j})^2\right),
 	\label{eq:pol}
 \end{align}
where $j=A,B$ denotes the sublattices, and the tilt angles of the tetrahedra are  $\kappa_A=-\kappa_B=\kappa$. 
The transformation properties of the site indices, position vectors and the spins and polarizations are collected in Table~\ref{table:transformation properties}.

The irreducible representations of the $D_2(C_2)$ group together with the classification of the net and staggered spin and polarization components are collected in Table~\ref{table:B_zero_irreps}.  As expected, the $M^b$, $L^a$, and $\Pi^c$ are invariant, so even without an external magnetic field both canted antiferromagnetism and antiferro-polarization is allowed in domain I [see in Fig. 1(a) in the main text]. 
 
 The dynamical susceptibility responsible for the NDD is given by
\begin{equation}
  \chi^{em}_{uv}(\omega) \propto \sum_n \Re(\langle 0 | P^u |n \rangle \langle n | M^v | 0\rangle) \delta(\omega-E_n+E_0) 
    \label{eq:chiem}
\end{equation} 
where $u,v$ take the $a,b,c$ values, $ | 0\rangle$ is the ground state and the $ | n\rangle$'s are  excited states. 
We will get non-zero susceptibility if both $P^u $ and $M^v$ tranform according to the same irreducible representation. In Table~\ref{table:B_zero_matrix_elements} we have collected the possible combinations for different directions of light. It turns out that in zero field we may expect finite signal only for light propagation along $[010]$. We shall also note that since the Hamiltonian is almost $O(2)$ symmetric about the $[001]$ axis, the $M^c$ almost commutes with the Hamiltanion (it commutes with the most significant $J$, $J^c$ and $\Lambda$ terms), and the matrix element $ \langle n | M^c | 0\rangle$ is small.  The $\chi^{em}_{ac}(\omega)$ only gets contribution from the small in-plane anisotropy and therefore it is much smaller than  $\chi^{em}_{ca}(\omega)$.




\setlength{\extrarowheight}{2pt}
\begin{table}[b]
	\caption{\label{table:B_par_z_irreps} Character table and transformation properties of net and staggered spin and polarization components, when the external magnetic field points in the $z$ direction.}
	\begin{ruledtabular}
	\begin{tabular}{ c rr c }		
		Irrep	& $\mathbb{1}$ & $\Theta \tilde{C}_2^{a}$ & Operators \\ \hline
		$A_1$	& $1$ & $1$ & $M^b, M^c, L^a, P^a, \Pi^b, \Pi^c $ \\
		$A_2$	& $1$ & $-1$ & $M^a, L^b, L^c, \Pi^a, P^b, P^c$ \\  
	\end{tabular}
\end{ruledtabular}
\end{table}

\begin{table}[t]
	\caption{\label{table:B_par_z_elements} Magnetoelectric susceptibility $\chi^{em}(\omega)$ in finite external magnetic field $\| [001]$ in the domain I for different light directions and polarizations, based on the character table for $2'_1$. Each entry in the table contains two data: the type of the configuration (Faraday or Voigt), and the information about the existence (and magnitude) of the appropriate susceptibility component. Our experimental setup corresponds to the Voigt geometry set in boldface.}
	\begin{ruledtabular}
		\begin{tabular}{ ccccc}		
 && $\mathbf{H}^{\omega}\|[100]$ & $\mathbf{H}^{\omega}\|[010]$ & $\mathbf{H}^{\omega}\|[001]$ \\
 &irrep& $A_2$ & $A_1$ & $A_1$ \\
\hline
\multirow{2}*{$\mathbf{E}^{\omega}\|[100]$} & \multirow{2}*{$A_1$}   & \multirow{2}*{$\times$} & Faraday & \bf{Voigt} \\
&  &  & finite & finite (small) \\
\multirow{2}*{$\mathbf{E}^{\omega}\|[010]$} & \multirow{2}*{$A_2$}   & Faraday & \multirow{2}*{$\times$}  & Voigt \\
&  & finite &  & 0 \\
\multirow{2}*{$\mathbf{E}^{\omega}\|[001]$} & \multirow{2}*{$A_2$}   & \bf{Voigt}  & Voigt & \multirow{2}*{$\times$}  \\
&  & finite & 0 &  \\
		\end{tabular}
	\end{ruledtabular}
\end{table}

\subsection{Finite external magnetic field $\mathbf{H}\| [001]$}

If we apply an external field pointing in the $c$ direction,  only the 
 \begin{align}
	\left\{\mathbb{1},\{\Theta C_2^{a}|[\frac{1}{2}00]\} \right\}
	\cong
	\left\{\mathbb{1}, \Theta \tilde{C}_2^{a} \right\}
	\label{eq:C2C1}
\end{align}
generators remain. This magnetic point group is the $2'$ or $C_2(C_1)$, with the unitary group element being just the identity.   There are only two irreps, $A_1$ and $A_2$, and the classification of the uniform and staggered spin and polarization components is shown in Table \ref{table:B_par_z_irreps}. 

Noting that the time-reversal operator can be expressed as $\Theta=C_2^b\mathcal{K}$ in the diagonal $S^c$ basis, where $\mathcal{K}$ is the operator of complex conjugation, we can simplify the anti-unitary $\Theta \tilde{C}_2^{a}$ operator as follows:
\begin{equation}
 \Theta \tilde{C}_2^{a}
  = C_2^b\mathcal{K} \tilde{C}_2^{a} 
  = C_2^b (\tilde{C}_2^{a})^* \mathcal{K} 
  = C_2^b \tilde{C}_2^{a} \mathcal{K} 
  =  \tilde{C}_2^{c} \mathcal{K} 
\end{equation}
where we used that $(\tilde C_2^a)^* = \tilde C_2^a$ and we introduced the unitary  $\tilde C_2^{c} = C_2^b \tilde C_2^{a}$, with the property $(\tilde C_2^c)^{-1} = \tilde C_2^c$.
The transformation $\Theta \tilde C_2^a \{\mathcal{O}\}$ for matrices is now $(\tilde{C}_2^c \mathcal{K}) \mathcal{O}( \tilde C_2^c \mathcal{K})^{-1}$ and we can write 
\begin{equation}
(\tilde{C}_2^c \mathcal{K}){\mathcal{O}}( \tilde C_2^c \mathcal{K})^{-1}
= \tilde{C}_2^c \mathcal{K} {\mathcal{O}} (\tilde C_2^c)^{-1*} \mathcal{K}
= \tilde{C}_2^c {\mathcal{O}}^* \tilde C_2^c \;.
\label{eq:C2cOC2c}
\end{equation}

When we consider the effect of the 
$
 \Theta \tilde C_2^a \{\mathcal{O}\} = \pm \mathcal{O}
$
  symmetry on the matrix elements of physical observables, we find that all representations are even or odd, following Table~\ref{table:B_par_z_irreps}. In matrix representation, using Eq.~(\ref{eq:C2cOC2c}):
\begin{equation}
 \tilde{C}_2^c {\mathcal{O}}^* \tilde C_2^c=\pm \mathcal{O}.
\label{eq:UKOUKmatrix}
\end{equation}

Since the Hamiltonian (see Eq.~3 in the main text) is invariant under the action of $\Theta \tilde C_2^a$, $ \tilde{C}_2^c \mathcal{H}^* \tilde{C}_2^c = \mathcal{H}$, we may learn about the action of $\Theta \tilde C_2^a$ on the eigenstates $|n\rangle$ of the Hamiltonian
$ \mathcal{H}|n\rangle = E_n |n\rangle $.
Applying the symmetry,
\begin{align}
 \tilde{C}_2^c \mathcal{H}^* \tilde{C}_2^c |n\rangle = E_n |n\rangle, \\
  \mathcal{H}^* \tilde{C}_2^c |n\rangle = E_n \tilde{C}_2^c |n\rangle.
\end{align}
This means that for non-degenerate eigenvalues 
\begin{equation}
  \tilde{C}_2^c |n\rangle = |n\rangle^* 
  \label{eq:C2cn}
\end{equation}
  up to an irrelevant phase. 

Let us apply Eq.~(\ref{eq:UKOUKmatrix}) to the operators in the $\langle 0 |  P^u  |n \rangle \langle n | M^v | 0\rangle$ matrix elements for the dynamical susceptibility, Eq.~(\ref{eq:chiem}): 
\begin{align} 
 \langle 0 | \tilde{C}_2^c (P^u)^* \tilde{C}_2^c |n \rangle \langle n | \tilde{C}_2^c (M^v)^* \tilde{C}_2^c | 0\rangle 
 \nonumber\\
=  \langle 0 | (\pm P^u)  |n \rangle \langle n | (\pm M^v)  | 0\rangle  
\end{align} 
Using the transformation properties of the eigenstates given by Eq.~(\ref{eq:C2cn}), we get
\begin{align} 
\left(\langle 0 |  P^u  |n \rangle \langle n | M^v | 0\rangle \right)^*
&= (\pm )_{P^u}(\pm )_{M^v} \langle 0 | P^u  |n \rangle \langle n |  M^v  | 0\rangle \;.
\end{align} 
This restricts the matrix elements of the $\chi^{em}_{uv}(\omega)$ to be real when both $P^u$ and $M^v$ belong to the same irreducible representation of the $2'$ magnetic point group, and to be pure imaginary when they belong to different irreducible representations -- in this case we do not expect NDD signal.
These selection rules are collected in Table \ref{table:B_par_z_elements}.

\end{document}